\documentclass[12pt]{article}
\usepackage{a4wide}
\usepackage{amssymb}
\usepackage{graphicx}
\begin{document} {\renewcommand{\thefootnote}{\fnsymbol{footnote}}
  \hfill  IGPG--07/3--5, NSF--KITP--07--55\\
  \medskip
  \hfill gr--qc/0703144\\
  \medskip \begin{center}
    {\LARGE  Dynamical coherent states and physical solutions of quantum cosmological bounces}\\
    \vspace{1.5em} Martin Bojowald\footnote{e-mail address: {\tt
        bojowald@gravity.psu.edu}}
% and ... 
\\
\vspace{0.5em}
Institute for Gravitational Physics and Geometry,
The Pennsylvania State
University,\\
104 Davey Lab, University Park, PA 16802, USA\\
\vspace{0.5em}
and\\
\vspace{0.5em}
Kavli Institute for Theoretical Physics, University of California,\\
 Santa Barbara, CA 93106, USA\\
\vspace{1.5em}
\end{center}
}

\setcounter{footnote}{0}

\newtheorem{theo}{Theorem}
\newtheorem{lemma}{Lemma}
\newtheorem{defi}{Definition}

\newcommand{\proofend}{\raisebox{1.3mm}{\fbox{\begin{minipage}[b][0cm][b]{0cm}
\end{minipage}}}}
\newenvironment{proof}{\noindent{\it Proof:} }{\mbox{}\hfill \proofend\\\mbox{}}
\newenvironment{ex}{\noindent{\it Example:} }{\medskip}
\newenvironment{rem}{\noindent{\it Remark:} }{\medskip}

\newcommand{\case}[2]{{\textstyle \frac{#1}{#2}}}
\newcommand{\lP}{\ell_{\mathrm P}}

\newcommand{\md}{{\mathrm{d}}}
\newcommand{\tr}{\mathop{\mathrm{tr}}}
\newcommand{\sgn}{\mathop{\mathrm{sgn}}}

\newcommand*{\R}{{\mathbb R}}
\newcommand*{\N}{{\mathbb N}}
\newcommand*{\Z}{{\mathbb Z}}
\newcommand*{\Q}{{\mathbb Q}}
\newcommand*{\C}{{\mathbb C}}

\begin{abstract}
  A new model is studied which describes the quantum behavior of
  transitions through an isotropic quantum cosmological bounce in loop
  quantum cosmology sourced by a free and massless scalar field. As an
  exactly solvable model even at the quantum level, it illustrates
  properties of dynamical coherent states and provides the basis for a
  systematic perturbation theory of loop quantum gravity. The detailed
  analysis is remarkably different from what is known for harmonic
  oscillator coherent states. Results are evaluated with regard to
  their implications in cosmology, including a demonstration that in
  general quantum fluctuations before and after the bounce are
  unrelated. Thus, even within this solvable model the condition of
  classicality at late times does not imply classicality at early
  times before the bounce without further assumptions. Nevertheless,
  the quantum state does evolve deterministically through the bounce.
\end{abstract}

\section{Introduction}

An understanding of high curvature regimes of a universe is likely to
require a quantization of gravity which is non-perturbative and
background independent. Background independence means that one does
not base the theory on pre-existing causal or geometrical structures
because they are to be provided by the quantized gravitational field
itself. Not surprisingly, many difficulties in this new setting have
to be overcome even when the aim is ``only'' to verify that a proposed
theory will have the correct semiclassical limit. One first has to
determine appropriate semiclassical states of the {\em interacting}
quantum theory of gravity. Thus, already in the definition of states
in which the classical limit is to be probed one has to face quantum
dynamics.  Unlike perturbative quantizations on a background, no
exactly known free vacuum state is available which one could use to
determine properties of an interacting vacuum state perturbatively.
Vacuum or coherent states of general interacting theories then have to
be constructed anew, which can show properties quite different from
the well-known (Gaussian) states of free theories or the harmonic
oscillator. Although it is often assumed, Gaussian states may not
capture the right semiclassical properties in any given system.  They
may be assumed as ``prepared'' initial states, but crucial deviations
from Gaussian form can occur especially in systems with long evolution
times, for which cosmology is the example {\em par excellence.}

Quantum gravity is not only an interacting quantum field theory whose
interacting semiclassical states are to be determined, it also, in
general, has no close relation to a free quantum field theory as it is
often exploited in effective field theories of particle physics. For
correct predictions it is, first of all, necessary to determine
precise states which capture semiclassical properties. In this paper,
a model, introduced in \cite{BouncePert}, is studied which is exactly
solvable and includes characteristic effects of loop quantum gravity,
one candidate for a background independent quantization
\cite{Rov,ALRev,ThomasRev}. The model itself is based on loop quantum
cosmology \cite{LivRev}. With new techniques
\cite{EffAc,EffectiveEOM}, coherent state properties can be determined
explicitly. In this sense, the model is analogous to the harmonic
oscillator in quantum mechanics and it has indeed the same solvability
properties as explained in more detail below.  This will allow us to
perform a complete dynamical coherent state analysis, demonstrating
how properties can differ considerably for distinct systems even when
one considers only solvable models. The model we study is not only
illuminating in this regard, but it also is of direct physical
interest since it describes non-singular cosmological bounce models.

Bouncing solutions of cosmological models have recently received much
attention. Although they are generally very special, they can indicate
how transitions through the classical big bang singularity may be
possible. Many different examples exist by now, which have most
systematically been developed in loop quantum cosmology. Most
arguments are based on ``effective'' equations which import some
quantum effects into classical equations, and which sometimes allow
exact analytical solutions (such as in
\cite{KasnerBounce,ExactBounce,BouncePert}) or can at least be studied
numerically; see e.g.\
\cite{BounceClosed,Oscill,BounceQualitative,GenericBounce,svv}. Also
in this context the above question of what a semiclassical state of an
interacting quantum theory looks like is relevant, although it is
often overlooked. It enters in the derivation or justification of
those effective equations which are supposed to capture properties of
semiclassical states. If the correct type of semiclassical states is
not known, one cannot be sure to have included all relevant
corrections to the classical equations in the right way. As a
byproduct, our solvable model presents the first case of a complete
set of effective equations in quantum cosmology.

Our model, used to illustrate semiclassical state issues, is solvable
exactly at the quantum level \cite{BouncePert}.  This is much stronger
than having analytical solutions of effective equations since full
states, including not only expectation values but also fluctuations
and higher moments, are under full control. In this sense, the system
is comparable to a harmonic oscillator. A complete analysis becomes
possible, including e.g.\ the evolution of coherent states
and their long-term dynamical properties. This is especially relevant
in the light of recent numerical analyses of related models where
initial Gaussian states without squeezing were evolved \cite{APS}. We
will see that the coherent state structure of the model is much richer
than that of unsqueezed Gaussian states, with squeezing influencing
the general behavior significantly. This is an instructive example for
the importance of a dedicated coherent state analysis, rather than
taking over harmonic oscillator properties to a new model.

The model is paradigmatic for background independent quantum gravity
obtained from a loop quantization where the usual free field theory
basis is not available. Since the bounce model is exactly solvable, it
can provide a perturbative basis for quantum gravity including all
possible interactions and degrees of freedom. Thus, the form of
coherent states determined here is relevant not only for the model
itself but for quantum gravity in general. At this stage, perturbative
inhomogeneities are not included explicitly and thus the question of
how they evolve through a bounce is not addressed in this paper. We
rather show and emphasize that even the unperturbed isotropic
situation poses several important questions for how quantum
fluctuations of the isotropic mode evolve through a bounce. We follow
a general method, summarized in Sec.~\ref{Method}. The solvable models
relevant for cosmology are introduced and analyzed in
Sec.~\ref{Models}, and discussed more broadly in Sec.~\ref{Concl}.

\section{The method}
\label{Method}

In what follows, we will not use a fixed, or any, representation of
our quantum system on a specific Hilbert space. Rather, we take an
algebraic viewpoint and treat the algebra of basic operators, such as
$[\hat{q},\hat{p}]=i\hbar$ in quantum mechanics, together with the
Hamiltonian $\hat{H}$ as primary. The quantities we are most
interested in are expectation values $\langle\hat{q}\rangle_{\psi}=
\langle\hat{q}\rangle$, $\langle\hat{p}\rangle_{\psi}=
\langle\hat{p}\rangle$ in a given state $\psi$, which we often drop as
a label if no confusion can arise, and fluctuations and correlations
$\Delta q= \sqrt{\langle\hat{q}^2\rangle- \langle\hat{q}\rangle^2}$,
$\Delta p= \sqrt{\langle\hat{p}^2\rangle- \langle\hat{p}\rangle^2}$
and $C_{qp}= \frac{1}{2}\langle\hat{q}\hat{p}+\hat{p}\hat{q}\rangle-
\langle\hat{q}\rangle\langle\hat{p}\rangle$. Higher moments, involving
higher than quadratic powers of basic operators, could be included by
the same means although we will not need to do so here. Nevertheless,
it is important to note that, would we determine all the moments, we
could reconstruct the state $\psi$ provided that the moments satisfy
appropriate conditions. The most basic such condition is Heisenberg's
uncertainty relation $(\Delta q)^2(\Delta p)^2-C_{qp}^2\geq \hbar^2/4$
(also called Schr\"odinger--Robertson uncertainty relation in this
form) which will be used frequently below. As in \cite{EffAc}, we call
all fluctuations, correlations and higher moments {\em quantum
variables} since, unlike expectation values, they describe typical
quantum properties.

Determining the evolution of moments is thus sufficient to find
properties of states. One can side-step the explicit construction of
states in a representation by deriving and solving equations of motion
for moments directly, such as $\md \langle\hat{p}\rangle/\md t=
\langle[\hat{p},\hat{H}]\rangle/i\hbar$ and $\md (\Delta p)^2/\md t=
\langle[\hat{p}^2,\hat{H}]\rangle/i\hbar- 2\langle\hat{p}\rangle\md
\langle\hat{p}\rangle/\md t$. In general, this set of equations is
highly coupled because, unless $[\hat{p},\hat{H}]$ is linear in basic
operators, $\langle[\hat{p},\hat{H}]\rangle$ is a function of
expectation values and quantum variables. The quantum variables, in
turn, will satisfy equations of motion involving moments of higher
degree. (See \cite{EffAc,Karpacz} for examples.) This coupling
describes the back-reaction of spreading and deformations on the peak
trajectory of a state, which is a crucial quantum effect. It is, for
instance, the reason for the usual non-locality in time of effective
actions.

This is the place where solvability properties of a model become
important. If $\hat{H}$ is quadratic, for instance,
$[\hat{p},\hat{H}]$ and $[\hat{q},\hat{H}]$ will be linear in
$\hat{q}$ and $\hat{p}$ and $\langle\hat{q}\rangle$ and
$\langle\hat{p}\rangle$ will not couple to quantum variables. As is
well-known for the harmonic oscillator, the spreading of states then
does not influence the peak motion at all and no non-trivial quantum
corrections arise in effective equations. More generally, this happens
whenever basic operators taken together with the Hamiltonian form a
linear commutator algebra. Our solvable bounce model is precisely of
this type. 

It is then feasible to solve equations of motion for
$\langle\hat{q}\rangle$, $\langle\hat{p}\rangle$ and any desired
quantum variables directly without taking the detour of first
computing a state in a chosen representation. Many representation
dependent difficulties can be avoided, such as explicit formulas for
inner products and normalizations.\footnote{Still, the existence proof
  of a representation poses an important and sometimes challenging
  question. An explicit representation may also be necessary for
  foundational issues such as the measurement problem. Nonetheless,
  dynamics can be analyzed without all details of the representation,
  and is thus useful to keep separate from foundational issues which
  are not fully understood even for quantum mechanics, let alone
  quantum gravity.} Instead, properties of the Hilbert space
structure, such as the self-adjointness of operators, can be
implemented straightforwardly through reality conditions for the
solutions of $\langle\hat{q}\rangle$, $\langle\hat{p}\rangle$ and
quantum variables.

\section{A solvable bounce model and its properties}
\label{Models}

A free isotropic scalar field $\phi$ couples to gravity through the
Friedmann equation
\begin{equation} \label{Friedmann}
 \frac{3}{8\pi G}c^2\sqrt{p}=\frac{1}{2}p^{-3/2}p_{\phi}^2\,.
\end{equation}
We use canonical variables, explained in more detail in
\cite{IsoCosmo}, whose relation to the scale factor $a$, the scalar
$\phi$ and their derivatives in proper time $\tau$ is $c=\md
a/\md\tau$ (extrinsic curvature), $p^{3/2}=a^3$ (volume) and
$p_{\phi}= p^{3/2} \md\phi/\md \tau$. Geometrically, $p$ is the
isotropic component of a densitized triad and can be positive or
negative for the two triad representations. In what follows, we assume
positive $p$ without loss of generality and drop absolute values.
Moreover, for later convenience we rescale $\phi$ by $\sqrt{2}$ and
drop factors of $8\pi G/3$ such that $\{c,p\}=1$. Solving
(\ref{Friedmann}) for $p_{\phi}$, which is a constant of motion,
yields $p_{\varphi}\propto
\pm cp$, allowing four possible choices for the signs: $c$ can be
positive (expanding universe) or negative (contracting universe), and
for each case $p_{\phi}$ can take any sign (such that $\phi$ runs
opposite or along coordinate time $\tau$).  We can interpret $H=cp$ as
the Hamiltonian which generates the flow in the variable $\phi$,
playing the role of internal time. The deparameterized Hamiltonian
constraint then reads $p_{\phi}+H=0$.

This Hamiltonian is quadratic, although not of the harmonic oscillator
form, and thus solvable as explained above.  Unlike for the harmonic
oscillator, the Hamiltonian does not have a definite sign. One can
easily understand the behavior of solutions, and of the energy
spectrum of the quantum theory, by performing a canonical
transformation to new canonical variables $(\pi,q)$ in which
$c=\frac{1}{\sqrt{2}}(\pi+q)$, $p=\frac{1}{\sqrt{2}}(\pi-q)$. The
Hamiltonian then becomes the upside-down harmonic oscillator
$H=\frac{1}{2}\pi^2-\frac{1}{2}q^2$ which obviously allows positive as
well as negative energy solutions. Classical solutions can easily be
determined as $\pi(\phi)= A\cosh \phi+B\sinh \phi$, $q(\phi)=
B\cosh\phi+A \sinh\phi$. In terms of the integration constants $A$ and
$B$, the Hamiltonian is $H=\frac{1}{2}(A^2-B^2)$. Corresponding
solutions in the original variables are $c(\phi)=
\frac{1}{\sqrt{2}}(A+B) e^{\phi}$, $p(\phi)= \frac{1}{\sqrt{2}} (A-B)
e^{-\phi}$.

\begin{figure}
\begin{center}
  \includegraphics[height=.3\textheight]{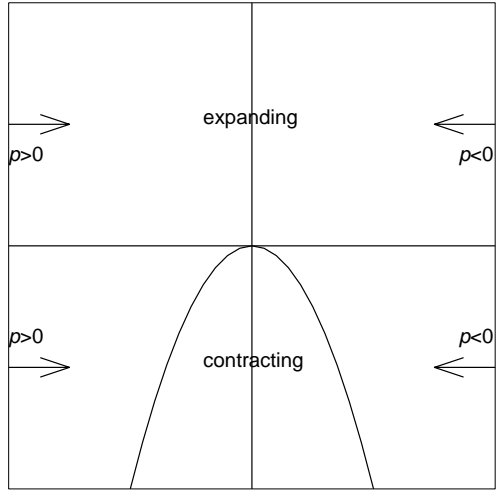}
  \caption{Illustration of positive and negative energy solutions in
  an upside-down harmonic oscillator potential
  $V(q)=-\frac{1}{2}q^2$. The sign of the corresponding triad variable
  $p$ is indicated.
\label{UpsideDown}}
\end{center}
\end{figure}

Since we assume positive $p$, solutions as functions of $q$ can only
be incoming from the left of the upside-down potential, where $\pi$ is
positive and $q$ negative. This assumption implies $A-B>0$. The sign
of $H$ then depends on whether we describe an expanding or contracting
universe, $c>0$ implying $H=\frac{1}{2}(A+B)(A-B)>0$ while $c<0$
implies $H<0$. In the first case, $p_{\phi}<0$ and $\phi$ runs
opposite to proper time, while it runs along proper time for a
collapsing universe. The opposite case would be realized had we chosen
the opposite sign for $H$. The behavior of solutions is illustrated in
Fig.~\ref{UpsideDown}.

This Hamiltonian is directly quantized, following the rules of quantum
mechanics, in the Wheeler--DeWitt approach \cite{DeWitt,QCReview}. It
is thus helpful to recall the properties of the resulting quantum
system as derived in \cite{BouncePert} using the method summarized in
Sec.~\ref{Method}. The quadratic Hamiltonian implies that the system
is solvable, which can be exploited to determine explicit solutions
$\langle\hat{c}\rangle(\phi)=c_1e^{\phi}$ and
$\langle\hat{p}\rangle(\phi)=c_2e^{-\phi}$ for expectation values of
the basic operators $\hat{c}$ and $\hat{p}$, in full agreement with
the classical solutions. Although the functional expressions for
expectation values agree with classical solutions, a difference to the
classical case is that states may be superpositions of positive and
negative eigenfunctions of $\hat{H}$ even though the expectation value
of $\hat{p}$ is either contracting or expanding. We will later discuss
the effect of possible admixtures of negative to positive energy
states in solutions.

Moreover, fluctuations can be solved explicitly by the same methods.
They are $(\Delta c)(\phi)^2= c_3 e^{2\phi}$, $(\Delta p)(\phi)^2 =
c_5 e^{-2\phi}$ and $C_{qp}(\phi)= c_4$ for the correlation. While
fluctuations are not constant, they satisfy $(\Delta
c)/\langle\hat{c}\rangle={\rm const}$ and $(\Delta
p)/\langle\hat{p}\rangle = {\rm const}$. Uncertainty relations require
$c_3c_5-c_4^2\geq \frac{1}{4}\hbar^2$. Typically, $\Delta c$ and
$\Delta p$ are thus of the order $\sqrt{\hbar}$, although the precise
value requires more specific information about the state. For
dimensional reasons, this must involve the only available scale $H$
such as in the form $\Delta p=\sqrt{\hbar H}$ and $\Delta c =
\sqrt{\hbar/H}$. In contrast to the harmonic oscillator where the
ground state provides specific values for fluctuations, no
distinguished state is available for the Hamiltonian encountered here.
Later on we will see that the loop quantization does provide more
information which one can use to estimate fluctuations. Still, the
information will not be as complete as it is for the harmonic
oscillator ground state.

The key feature of a loop quantization is that its representation does
not provide a $c$-operator but only operators for almost periodic
functions of $c$ \cite{Bohr} (i.e.\ countable superpositions of
$\exp(i\mu c)$ with $\mu\in\R$). This is sufficient for constructing a
Hamiltonian operator which is well-defined and agrees with the
classical one, $cp$, in low curvature regimes where $c\ll 1$. A loop
quantization can thus be understood as implying that $c$ in the
classical Hamiltonian is replaced by $\sin c$ when it is
quantized. Instead of $\sin c$ one could, of course, choose any other
almost periodic function which reduces to $c$ when $c\ll1$.  The
freedom can be restricted by relating the model to quantizations of
the Hamiltonian constraint as it can be defined in the full theory
\cite{QSDI}. This distinguishes the use of $\sin c$ as it follows from
the original derivations in
\cite{IsoCosmo}, although other choices remain \cite{AmbigConstr}. In
any case, the resulting Hamiltonian is no longer quadratic in the
variables $(c,p)$ and thus appears quite complicated.  However, if we
introduce the operators $\hat{p}$ and
$\hat{J}:=\hat{p}\widehat{\exp(ic)}$, the {\em linear} Hamiltonian
operator $\hat{H}=-\frac{1}{2}i(\hat{J}-\hat{J}^{\dagger})$ reduces to
the classical Hamiltonian when curvature $c$ is small, and it depends
on $c$ only through $\exp(ic)$ which is required by loop quantum
cosmology.  It can be viewed as a quantization of $p\sin c$ in a
specific factor ordering.

A linear Hamiltonian usually simplifies the dynamics very much, but
only if the basic variables in which it is linear form a closed
algebra (see \cite{EffAc} for a discussion in the general context of
effective equations). Since we transformed from canonical variables
$(c,p)$ to non-canonical ones $(p,J)$, the system is not guaranteed to
simplify even with a linear $\hat{H}$.  Fortunately, as one can easily
check the variables do satisfy the closed ${\rm sl}(2,\R)$ algebra
\begin{equation} \label{comm}
  [\hat{p},\hat{J}]=\hbar\hat{J}\quad,\quad {}
  [\hat{p},\hat{J}^{\dagger}]=-\hbar\hat{J}^{\dagger}\quad,\quad {}
  [\hat{J},\hat{J}^{\dagger}]=-2\hbar\hat{p}-\hbar^2\,.
\end{equation}
which includes the Hamiltonian as a linear combination of $\hat{J}$
and $\hat{J}^{\dagger}$.\footnote{This algebra has repeatedly occurred
  in canonical quantum gravity, for instance to quantize black hole
  phase spaces \cite{GroupArea,Group,Orientation}. The analog of
  $\hat{p}$ in this case is the mass of the black hole, which one
  requires to be positive, selecting the discrete positive series of
  ${\rm sl}(2,\R)$-representations. Note the resemblence of the basic
  variables $p$ and $J$ to the variables used in affine quantum
  gravity \cite{AffineQG,AffineQG2}. In that case, as with $\hat{J}$
  here, one multiplies the metric with momenta to ensure the existence
  of representations of the resulting affine commutation relations
  where the metric is positive definite. Although we use triad
  variables and do not need to require positivity for them, there are
  representations where $\hat{p}$ is indeed positive definite. Also in
  quantum optics, and even for the harmonic oscillator itself, the
  ${\rm sl}(2,\R)$ algebra plays a role
  \cite{OpticsPhaseSpace,AngleJ,HarmOsc}.} This is the reason for the
exact solvability of the system which we will use below.
Representations of ${\rm sl}(2,\R)$ do not allow one to have a purely
positive spectrum for $i(\hat{J}-\hat{J}^{\dagger})$. Thus, any
representation space contains positive and negative energy solutions
and we will have to dicsuss appropriate restrictions on states to rule
out superpositions of negative energy contributions to positive energy
states if superpositions of expanding and collapsing universe branches
are not to be allowed.

Before analyzing equations of motion we note that the system can be
generalized by performing a canonical transformation such as
$\pi:=p^kc$, $v:=p^{1-k}/(1-k)$ and using the new canonical variables
$(\pi,v)$ instead of $(c,p)$ in the definition of $\hat{p}$ and
$\hat{J}$.  Properties of the system to be discussed below are not
affected by this re-interpretation of the variables. Such
transformations can be motivated by taking into account features of
nearly isotropic but inhomogeneous lattice states of loop quantum
gravity and possible dynamical refinements of the underlying lattice
\cite{InhomLattice,APSII}. The special value $k=0$ then corresponds to
a fixed lattice as realized in \cite{IsoCosmo,Bohr}, while $k=-1/2$,
introduced independently in \cite{APSII}, corresponds to a refinement
such that the number of lattice vertices increases linearly as a
function of volume.

\subsection{Equations of motion}

As discussed before, the system can much more easily be understood if
we do not first solve for wave functions, subject to
$i\hbar\dot{\psi}= \hat{H}\psi$, and then compute expectation values
and fluctuations from solutions. Instead, using a more algebraic point
of view we derive equations of motion directly for expectation values,
fluctuations and higher moments and solve them
\cite{EffAc,Karpacz}.

For expectation values, now simply denoted as
$p:=\langle\hat{p}\rangle$, $J:=\langle\hat{J}\rangle$ and
$\bar{J}:=\langle\hat{J}^{\dagger}\rangle$ in an arbitrary normalized
state, equations of motion follow immediately by taking expectation
values of Heisenberg equations of motion, or by using the
Schr\"odinger equation for the state appearing in the expectation value,
\begin{equation}
 \dot{p}= \frac{1}{i\hbar}\langle[\hat{p},\hat{H}]\rangle=
 -\frac{1}{2}(J+\bar{J}) \quad,\quad 
 \dot{J}= \frac{1}{i\hbar}\langle[\hat{J},\hat{H}]\rangle=
 -p-\frac{1}{2}\hbar=\dot{\bar{J}}\,. \label{Jdot}
\end{equation}
That these equations form a closed system is a consequence of the
linear nature of the variables and Hamiltonian. In general, the
evolution of expectation values would also depend on all higher
moments of the state: During evolution the state spreads and deforms,
which then back-reacts on the peak position of a wave packet. This
back-reaction is the dynamical essence of a quantum system, captured
in effective equations.

In fact, a state is characterized not just by its expectation values
of basic operators but also by the infinitely many quantum variables
which specify fluctuations and higher moments of a state. Fluctuations
(and correlations) can be defined, as usually, by
\begin{eqnarray}
 G^{pp} &:=& \langle\hat{p}^2\rangle-p^2\\
 G^{JJ} &:=& \langle\hat{J}^2\rangle-J^2\quad,\quad
 G^{\bar{J}\bar{J}} :=
 \langle\hat{J}^{\dagger}{}^2\rangle-\bar{J}^2\\
 G^{pJ} &:=& \frac{1}{2}\langle
 \hat{p}\hat{J}+\hat{J}\hat{p}\rangle- pJ\quad,\quad
 G^{p\bar{J}} := \frac{1}{2}\langle
 \hat{p}\hat{J}^{\dagger}+\hat{J}^{\dagger}\hat{p}\rangle- p\bar{J}\\
  G^{J\bar{J}} &:=& \frac{1}{2}\langle
 \hat{J}\hat{J}^{\dagger}+\hat{J}^{\dagger}\hat{J}\rangle- |J|^2\,.
\end{eqnarray}
Since we use partially complex variables there are initially more than
three independent fluctuations. However, reality conditions to be
imposed later at the quantum level will lead to additional relations
and reduce the number of independent degrees of freedom to the correct
value. Higher moments are defined analogously, using totally symmetric
orderings in expressions where both $\hat{p}$ and $\hat{J}$ are involved.

Fluctuations are not expectation values of a single operator and their
equations of motion do not follow directly as before. But they can
easily be derived using linearity and the Leibniz rule. We have,
e.g.,
\begin{eqnarray}
 \dot{G}^{pp} &=& \frac{1}{i\hbar} \langle[\hat{p}^2,\hat{H}]\rangle-
 2p\dot{p}
 = -\frac{1}{2}\langle \hat{p}\hat{J}+\hat{J}\hat{p}+
 \hat{p}\hat{J}^{\dagger}+ \hat{J}^{\dagger}\hat{p}\rangle+
 p(J+\bar{J})\\
 &=& -G^{pJ}-G^{p\bar{J}}
\end{eqnarray}
and similarly
\begin{eqnarray}
 \dot{G}^{JJ} &=& -2G^{pJ}\quad,\quad
 \dot{G}^{\bar{J}\bar{J}} = -2G^{p\bar{J}}\\
 \dot{G}^{pJ} &=& -\frac{1}{2} G^{JJ}-\frac{1}{2}G^{J\bar{J}}-G^{pp} 
\quad,\quad
 \dot{G}^{p\bar{J}} = -\frac{1}{2} G^{\bar{J}\bar{J}}-
  \frac{1}{2}G^{J\bar{J}}-G^{pp}\\
 \dot{G}^{J\bar{J}} &=& -G^{pJ}-G^{p\bar{J}}\,.
\end{eqnarray}
Higher moments also are subject to equations of
motion which follow analogously.

For our solvable system, all these equations of motion are linear in
the quantum variables and only finitely many ones are coupled to each
other. They can thus be solved straightforwardly, such as
\begin{eqnarray}
  p(\phi) &=& {\textstyle\frac{1}{2}}(Ae^{-\phi}+Be^{\phi})-
 {\textstyle\frac{1}{2}}\hbar\,, \label{psol}\\
 J(\phi) &=& {\textstyle\frac{1}{2}}(Ae^{-\phi}-Be^{\phi})+iH \label{Jsol}
\end{eqnarray}
(using $-\frac{1}{2}i(J-\bar{J})=H:=\langle\hat{H}\rangle$) for the
expectation values. To decouple the six equations for fluctuations, we
first note that $\dot{G}^{J\bar{J}}=\dot{G}^{pp}$ and
$\frac{1}{2}(\dot{G}^{JJ}+\dot{G}^{\bar{J}\bar{J}})= \dot{G}^{pp}$
implies
\begin{eqnarray}
 G^{J\bar{J}}-G^{pp}&=&c_1 \label{EOMc1}\\
 \frac{1}{2}G^{JJ}+\frac{1}{2}G^{\bar{J}\bar{J}}- G^{pp} &=& c_2
\end{eqnarray}
with constants $c_1$ and $c_2$. Moreover, only two sets of two coupled
equations, one for $G^{pp}$ and $G^{pJ}+G^{p\bar{J}}$,
\begin{eqnarray*}
 \dot{G}^{pp} &=& -G^{pJ}-G^{p\bar{J}}\\
 \frac{\md}{\md t}(G^{pJ}+G^{p\bar{J}}) &=& -c_1-c_2-4G^{pp}
\end{eqnarray*}
 and one for
$G^{pJ}-G^{p\bar{J}}$ and $G^{JJ}-G^{\bar{J}\bar{J}}$,
\begin{eqnarray*}
 \frac{\md}{\md t}(G^{pJ}-G^{p\bar{J}}) &=&
 -\frac{1}{2}(G^{JJ}-G^{\bar{J}\bar{J}})\\
 \frac{\md}{\md t}(G^{JJ}-G^{\bar{J}\bar{J}}) &=&
 -2(G^{pJ}-G^{p\bar{J}})
\end{eqnarray*}
remain. They yield
\begin{eqnarray}
 G^{pp}(\phi) &=& \frac{1}{2}(c_3e^{-2\phi}+c_4e^{2\phi})- \frac{1}{4}(c_1+c_2)\\
 G^{JJ}(\phi) &=& \frac{1}{2}(c_3e^{-2\phi}+c_4e^{2\phi})+\frac{1}{4}(3c_2-c_1)-
 i(c_5e^{\phi}-c_6e^{-\phi})\\
 G^{\bar{J}\bar{J}} &=& \frac{1}{2}(c_3e^{-2\phi}+c_4e^{2\phi})+\frac{1}{4}(3c_2-c_1)+
 i(c_5e^{\phi}-c_6e^{-\phi})\\
 G^{pJ}(\phi) &=& \frac{1}{2}(c_3e^{-2\phi}-c_4e^{2\phi})+ \frac{i}{2}(c_5e^{\phi}+c_6e^{-\phi})\\
 G^{p\bar{J}}(\phi) &=& \frac{1}{2}(c_3e^{-2\phi}-c_4e^{2\phi})- \frac{i}{2}(c_5e^{\phi}+c_6e^{-\phi})\\
 G^{J\bar{J}}(\phi) &=& \frac{1}{2}(c_3e^{-2\phi}+c_4e^{2\phi})+
 \frac{1}{4}(3c_1-c_2)
\end{eqnarray}
with further constants of integration.

\subsection{Physical inner product properties}

Although we are not dealing explicitly with states, properties of the
Hilbert space must be reflected in the structure of our quantum
variables. The equations of motion, for instance, are only valid if we
understand the quantum variables to be defined with normalized states
since $\langle|\rangle=1$ has been used, for instance when taking an
expectation value of $[\hat{J},\hat{J}^{\dagger}]=
-2\hbar\hat{p}-\hbar^2$ to compute $\dot{J}$ in (\ref{Jdot}). The
system is, initially, defined through the Friedmann equation as a
constraint on the space where both the metric and scalar field degrees
of freedom are quantized. After quantizing this equation, it is
re-interpreted as describing relational motion in $\phi$. Loop quantum
cosmology provides the kinematical inner product on the original
Hilbert space where the constraint operator is defined, but not
immediately one on the solution space.  Such a physical inner product
can be difficult to determine explicitly in a representation of
states. It can be derived, for instance, by requiring that basic
operators are self-adjoint if they correspond to real classical
variables. This then directly implies that all quantum variables
defined from self-adjoint operators must be real, which in our
procedure is the analog of using the physical inner product in
deriving expectation values through states. Since such reality
conditions can be imposed directly for quantum variables, implementing
physical inner product properties at this level can be much more
straightforward than at the level of states. One reason is the
representation independence of the formalism which allows one to avoid
looking for a representation of states in which a computation of the
physical inner product may be feasible. This is especially useful for
quantum gravity where the general physical inner product problem is
one of the major issues.

\subsubsection{Reality conditions}

In our case, we use one complex classical variable $J=pe^{ic}$, and
thus cannot refer to a self-adjoint quantization of $c$ directly since
no such operator exists at all in a loop quantization.  Reality
conditions implementing the physical inner product must be formulated
in a more complicated way: In addition to the simple adjointness
relation $\hat{p}^{\dagger}=\hat{p}$ quantizing the real variable $p$,
we have a non-linear relation
\begin{equation}
 \hat{J}\hat{J}^{\dagger}=\hat{p}^2
\end{equation}
which follows from the fact that $e^{ic}$ must be quantized to a
unitary operator for $c$ to be real. (However,
$\hat{J}^{\dagger}\hat{J}\not=\hat{p}^2$ in the ordering chosen for
the definition of $\hat{J}=\hat{p}\widehat{e^{ic}}$.)

Taking expectation values of this equation, we obtain a relation
between quantum variables and expectation values: using the
commutation relations (\ref{comm}) in
\begin{equation}
 G^{J\bar{J}}= \frac{1}{2}\langle\hat{J}\hat{J}^{\dagger}+
 \hat{J}^{\dagger}\hat{J}\rangle- |J|^2=
 \langle\hat{J}\hat{J}^{\dagger}\rangle+ \hbar p+\frac{1}{2}\hbar^2-|J|^2
\end{equation}
we have
\begin{eqnarray}
\langle\hat{J}\hat{J}^{\dagger}\rangle &=& G^{J\bar{J}}-\hbar
p-\frac{1}{2}\hbar^2+|J|^2\nonumber\\
 &=& \langle\hat{p}^2\rangle= G^{pp}+p^2
\end{eqnarray}
and thus
\begin{equation} \label{reality}
 |J|^2-(p+{\textstyle\frac{1}{2}}\hbar)^2=G^{pp}-G^{J\bar{J}}+
 \frac{1}{4}\hbar^2=
 \frac{1}{4}\hbar^2-c_1\,.
\end{equation}
This condition mixes expectation values and quantum variables, but
depends on fluctuations only through the constant $c_1$. An initial
state thus determines how the reality conditions between expectation
values are realized. The relation (\ref{reality}) is then preserved in
time since $G^{J\bar{J}}-G^{pp}=c_1$ is constant as derived in
(\ref{EOMc1}). If fluctuations are small, as in semiclassical states,
we have $|J|^2=p^2+O(\hbar)$ which, as it should, is the classical
relation satisfied up to quantum corrections.

\subsubsection{Sign of the energy}

Although not directly related to the physical inner product, we
include in this section a discussion of the requirement of a definite
sign for the Hamiltonian in superposed states. One may or may not wish
to allow superpositions of expanding and contracting universe branches
in quantum cosmology, but arguments using the physical inner product
in analogy to the Klein--Gordon equation suggest that only energy
eigenstates of a definite sign should be allowed in superpositions
\cite{APS}.  For a linear system, we did not explicitly take an
absolute value of the Hamiltonian operator, which implies that in
general we are not guaranteed that only positive energy solutions
enter states corresponding to our solutions. In a language more
suitable to quantum cosmology, ``positive energy'' means that at any
fixed time $\phi$ we should not allow superpositions of expanding and
contracting branches of a universe (while at different times the
universe certainly does not need to be always expanding or always
contracting).  To rule out significant contributions from a superposed
branch we have to pose further conditions for our quantum variables
ensuring that they arise from expectation values in states which are
superpositions of only positive energy eigenstates (or only negative
energy eigenstates). In general, expressing the positivity of
operators through expectation values can be complicated.

But for our purposes it is, fortunately, possible to proceed without
technical complications. We will be interested in states which at some
point (e.g.\ at late times) are semiclassical. This restricts the
values that fluctuations can take compared to the magnitude of
expectation values. It refers in particular to $p_{\phi}$ as one of
the matter variables. We require that its fluctuation is small
compared to its expectation value which, through the dynamical
equation, implies the relation
\begin{equation} \label{GHHsmall}
  G^{HH}:=\langle\hat{H}^2\rangle-\langle\hat{H}\rangle^2\ll
  \langle\hat{H}\rangle^2\,.
\end{equation}
If this is realized and $H=\langle\hat{H}\rangle>0$, a state in the
$H$-representation, i.e.\ written as a superposition of
$\hat{H}$-eigenstates, is sharply peaked at a large positive value of
$H$. Thus, there are no significant contributions from negative
energy states. (We will return to this issue in
Sec.~\ref{role}.) Since $H$ and $G^{HH}$ are constant during evolution,
imposing (\ref{GHHsmall}) at one initial time ensures that it is
satisfied at all times.

We can express this condition in terms of the integration constants
derived before. From
\begin{eqnarray}
 G^{HH} &=& \langle\hat{H}^2\rangle-H^2\\
 &=& -\frac{1}{4}\langle\hat{J}^2- \hat{J}\hat{J}^{\dagger}-
 \hat{J}^{\dagger}\hat{J}+ \hat{J}^{\dagger}{}^2\rangle+
 \frac{1}{4}(J^2-2|J|^2+\bar{J}^2)\\
 &=& -\frac{1}{4}(G^{JJ}+G^{\bar{J}\bar{J}})+\frac{1}{2}G^{J\bar{J}}=
 \frac{1}{2}(c_1-c_2)
\end{eqnarray}
we see that $c_1-c_2$ must be small compared to $H^2$. (But it cannot
be zero since $\hat{H}$ has continuous spectrum.) This is the primary
condition we have to impose not only for semiclassicality but also to
ensure that a state is dominated by contributions of definite energy
sign.  Later, we will add further semiclassicality conditions to
restrict also the fluctuations of other variables such as $p$.

\subsection{Bouncing solutions}

Our general solution (\ref{psol}) for $p$ allows bouncing\footnote{The
bounce occurs at $p\approx H$ which can be much larger than $\hbar$
although it is quantum effects which cause it. Deviations from
classical behavior occur in quantum geometry due to large curvature
which, for a large matter content, can be realized even when the
spatial volume is still large compared to the Planck length.}
solutions for $AB>0$ as well as ``singular'' solutions for $AB<0$
which reach $p=0$ in finite time $\phi$. (Although isotropic loop
quantum cosmology is non-singular for any solution \cite{Sing,BSCG},
additional correction terms become manifest at small volume which are
not included here in the solvable model. The model itself thus breaks
down before $p=0$ is reached. The singularity in our equations only
indicates that a deep quantum geometry regime is reached, just as one
commonly expects the general singularity problem to be resolved.
Nevertheless, we keep solutions with $AB<0$ for now since they will be
ruled out even within our model shortly.)  The internal time variable
$\phi$ has just been chosen for convenience of the mathematical
description, rather than referring to physical observers. For a
solution reaching $p=0$ to be considered singular one must also verify
that {\em proper} time remains finite. We thus need to interpret our
relational solution ($p$ and $J$ as functions of $\phi$) as a
space-time geometry subject to modified dynamics as it arises from the
loop quantization.

We do not have any manifold picture, except for the homogeneous
spatial manifold we started with to reduce the classical system. What
is missing is a manifold for the time extension, which is
indispensable if we want to compute a proper time interval. A time
direction and coordinate can be introduced by reverting back to the
Friedmann formulation of constrained dynamics.  We interpret the
effective Hamiltonian density
$p^{-3/2}\langle\hat{H}\rangle^2=p^{-3/2}({\rm
  Im}J)^2=\sqrt{p}\sin^2c$ together with the matter contribution
$\frac{1}{2}p^{-3/2}p_{\phi}^2$ as an effective constraint\footnote{We
  emphasize that this way to derive an effective constraint is correct
  only due to the linearity in $\hat{J}$ of the Hamiltonian $\hat{H}$.
  Non-linear terms in the Hamiltonian, as they arise from any
  deviation from the solvable model such as by including a matter
  potential, imply additional correction terms depending on quantum
  variables. They would have to be studied carefully for the
  derivation of effective equations of non-solvable systems; see e.g.\
  \cite{EffAc,EffectiveEOM,Karpacz} for the low energy effective
  action as an example.}
\[
 C=-\sqrt{p}\sin^2c+\frac{4\pi G}{3}p^{-3/2}p_{\phi}^2
\]
which generates coordinate evolution (in $\eta$) in a gauge specified
by the lapse function $N$, $\md p/\md \eta = \{p,NC\}$. For proper time,
$\eta=\tau$, we simply have $N=1$ and thus
\[
 \md p/\md \tau=\{p,C\}=\sqrt{p}\sin(2c)\,.
\]
From the equation
\begin{equation}
 \sin(2c(\phi))=\frac{1}{\sqrt{p(\phi)}}\frac{\md p(\phi)}{\md\tau}=
 \frac{1}{\sqrt{p(\phi)}} \frac{\md p}{\md\phi}
 \frac{\md\phi}{\md\tau}
\end{equation}
and our solutions (\ref{psol}) and (\ref{Jsol}) we can then compute
$\phi(\tau)$ by integrating
\begin{equation}
  \frac{\md\phi}{\md\tau}= \sqrt{p(\phi)}\frac{2\sin(c(\phi))\cos(c(\phi))}{\md
    p/\md\phi}= \frac{-2\sqrt{2}H}{(Ae^{-\phi}+Be^{\phi}-\hbar)^{3/2}}
\end{equation}
(using $J/p=\cos c+i\sin c$). We can always assume that either $A=B>0$
(for a bouncing solution) or $A=-B>0$ (for a non-bouncing one) since
we only need to shift the origin of $\phi$ if $|A|\not=|B|$. This
leaves us with two cases,
\begin{equation}
 \tau(\phi)= -\frac{A^{3/2}}{H}\int^{\phi}\cosh^{3/2}(z)\md z
\end{equation}
for $A=B$ and
\begin{equation}
 \tau(\phi)= -\frac{A^{3/2}}{H}\int^{\phi}\sinh^{3/2}(z)\md z
\end{equation}
for $A=-B$. The integrals can be determined in terms of elliptic
functions, but we are only interested in the fact that $\tau(\phi)$ is
finite at any finite value of $\phi$ which can be seen directly from
the integrals. Thus, proper time remains finite when $p=0$ is reached.

Singular solutions could thus be possible.  But not all these solutions
satisfy the reality condition (\ref{reality}) which still has to be
imposed. From its general form we obtain
\[
 |J|^2-(p+{\textstyle\frac{1}{2}}\hbar)^2= -AB+H^2=\frac{1}{4}\hbar^2-c_1
\]
and thus
\begin{equation} \label{realityAB}
 AB=H^2+c_1-\frac{1}{4}\hbar^2\,.
\end{equation}
For macroscopic values of $H$ and small (or positive) $c_1$ from
fluctuations, we only have bouncing solutions with $AB>0$. Singular
solutions can only be obtained for large and negative $c_1$ which is
never realized for states which are semiclassical at one time. Note,
however, that $c_1$ can be large even if our condition $G^{HH}\ll H^2$
which is necessary for solutions to respect positivity is realized
since the latter condition only constrains $c_1-c_2$.  Our discussion
thus shows that it is crucial to know and use the reality conditions,
or ultimately the physical inner product, to draw conclusions about
bouncing solutions versus non-bouncing ones.

\subsection{Uncertainty relations}

Fluctuations cannot take arbitrary values but are restricted by
uncertainty relations. For each pair of {\em self-adjoint} basic
operators we have one uncertainty relation, which in our case implies
three relations since $\hat{J}+\hat{J}^{\dagger}$ and
$i(\hat{J}-\hat{J}^{\dagger})$ are independent in addition to
$\hat{p}$. For each pair $(\hat{A},\hat{B})$ of self-adjoint
operators, as usually, we start from the Schwarz inequality
\[
 \langle\psi_1|\psi_1\rangle \langle\psi_2|\psi_2\rangle \geq
 |\langle\psi_1|\psi_2\rangle|^2
\]
applied to\footnote{$\widehat{\Delta A}$ is not a linear operator due
  to the dependence on the state in $\langle\hat{A}\rangle$, but
  $|\psi_1\rangle$ is well-defined as a state obtained from
  $|\psi\rangle$.} $|\psi_1\rangle:=\widehat{\Delta A}|\psi\rangle$
and $|\psi_2\rangle:=\widehat{\Delta B}|\psi\rangle$ with
$\widehat{\Delta A}:=\hat{A}- \langle\hat{A}\rangle$:
\begin{equation}
 \langle(\widehat{\Delta A})^2\rangle \langle(\widehat{\Delta
   B})^2\rangle \geq |\langle\widehat{\Delta A} \widehat{\Delta
   B}\rangle|^2\,.
\end{equation}
Writing
\[
 \widehat{\Delta A}\widehat{\Delta B}= \frac{1}{2}(\widehat{\Delta A}
 \widehat{\Delta B}+ \widehat{\Delta B} \widehat{\Delta A})+
 i\frac{1}{2i} [\widehat{\Delta A},\widehat{\Delta B}]
\]
and
\begin{eqnarray*}
 \frac{1}{2}\langle\widehat{\Delta A} \widehat{\Delta B}+
 \widehat{\Delta B} \widehat{\Delta A}\rangle &=& \frac{1}{2}\langle
 \hat{A}\hat{B}+\hat{B}\hat{A}\rangle- AB\\
\mbox{} [\widehat{\Delta A},\widehat{\Delta B}] &=& [\hat{A},\hat{B}]
\end{eqnarray*}
we have
\[
 |\langle\widehat{\Delta A} \widehat{\Delta
   B}\rangle|^2= \frac{1}{4}(\langle \hat{A}\hat{B}+ \hat{B}\hat{A}\rangle^2
 -2\langle\hat{A}\rangle\langle\hat{B}\rangle)^2+
 \frac{1}{4}\langle-i[\hat{A},\hat{B}]\rangle^2
\]
where we used self-adjointness of the operators to compute the
absolute square of the complex number $\langle\widehat{\Delta A}
\widehat{\Delta B}\rangle$. In terms of quantum variables, we thus
have the general form
\begin{equation}
 G^{AA}G^{BB}-(G^{AB})^2\geq \frac{1}{4}\langle
 -i[\hat{A},\hat{B}]\rangle^2
\end{equation}
of uncertainty relations whenever $\hat{A}$ and $\hat{B}$ are
self-adjoint.

Specifically, we have three pairs
$(\hat{p},\hat{J}+\hat{J}^{\dagger})$,
$(\hat{p},i(\hat{J}-\hat{J}^{\dagger}))$ and
$(\hat{J}+\hat{J}^{\dagger}, i(\hat{J}-\hat{J}^{\dagger}))$ of
different self-adjoint basic operators. We obtain three inequalities
involving $G^{pp}$ and the fluctuations
\begin{eqnarray}
 G^{J+\bar{J},J+\bar{J}} &:=&
 \langle(\hat{J}+\hat{J}^{\dagger})^2\rangle- (J+\bar{J})^2\nonumber\\
 &=& G^{JJ}+2G^{J\bar{J}}+G^{\bar{J}\bar{J}}= 4G^{pp}+2(c_1+c_2)\,,\\
 G^{i(J-\bar{J}),i(J-\bar{J})} &:=&
 -\langle(\hat{J}-\hat{J}^{\dagger})^2\rangle+ (J-\bar{J})^2\nonumber\\
 &=& 4G^{HH}= 2(c_1-c_2)\,,\\
 G^{p,J+\bar{J}} &:=&
 \frac{1}{2}\langle\hat{p}(\hat{J}+\hat{J}^{\dagger})+
 (\hat{J}+\hat{J}^{\dagger})\hat{p}\rangle- p(J+\bar{J})\nonumber\\
 &=& G^{pJ}+G^{p\bar{J}}\,,\\
 G^{p,i(J-\bar{J})} &:=& \frac{i}{2}\langle\hat{p}(\hat{J}-\hat{J}^{\dagger})+
 (\hat{J}-\hat{J}^{\dagger})\hat{p}\rangle- ip(J-\bar{J})\nonumber\\
 &=& i(G^{pJ}-G^{p\bar{J}})\,,\\
 G^{J+\bar{J},i(J-\bar{J})} &:=&
 \frac{i}{2}\langle(\hat{J}+\hat{J}^{\dagger})(\hat{J}-\hat{J}^{\dagger})+ (\hat{J}-\hat{J}^{\dagger})(\hat{J}+\hat{J}^{\dagger})\rangle- i(J+\bar{J})(J-\bar{J})\nonumber\\
 &=& i(G^{JJ}-G^{\bar{J}\bar{J}})\,.
\end{eqnarray}
Using the explicit solutions, they are
\begin{eqnarray}
 G^{pp}G^{J+\bar{J},J+\bar{J}}- (G^{p,J+\bar{J}})^2 &=&
 4c_3c_4-\frac{1}{4}(c_1+c_2)^2 \label{uncertI}\\
 &\geq& \hbar^2H^2\nonumber\\
 G^{pp}G^{i(J-\bar{J}),i(J-\bar{J})}- (G^{p,i(J-\bar{J})})^2 &=&
 (c_1-c_2) (c_3e^{-2\phi}+c_4e^{2\phi})+ \frac{1}{2}(c_2^2-c_1^2)\label{uncertII}\\
 &&-
 c_5^2e^{2\phi}- 2c_5c_6- c_6^2e^{-2\phi}\nonumber\\
 &\geq& \frac{1}{4}\hbar^2(J+\bar{J})^2=
 \frac{1}{4}\hbar^2(A^2e^{-2\phi}-2AB+B^2e^{2\phi})\nonumber\\
 G^{J+\bar{J},J+\bar{J}}G^{i(J-\bar{J}),i(J-\bar{J})}-
 (G^{J+\bar{J},i(J-\bar{J})})^2 &=& 4(c_1-c_2) (c_3e^{-2\phi}+c_4e^{2\phi})-
 2(c_2^2-c_1^2)\label{uncertIII}\\
&&- 4c_5^2e^{2\phi}+8c_5c_6- 4c_6^2e^{-2\phi}\nonumber\\
 &\geq& \hbar^2(2p+\hbar)^2= \hbar^2(A^2e^{-2\phi}+2AB+B^2e^{2\phi})\,.\nonumber
\end{eqnarray}

These uncertainty relations are the equations which determine
properties of semiclassical, near coherent states.  We will later
discuss these relations, in particular their saturation, in more
detail and give a complete analysis of coherent states of this
system. Before doing so we can already note here that there are quite
unfamiliar properties compared to what one knows from harmonic
oscillator coherent states. The first relation, (\ref{uncertI}) shows
that there is a type of uncertainty relation between the constants of
integration $c_3$ and $c_4$, one of which determines the
$p$-fluctuation before and one after the bounce. Thus, if the
uncertainties are very small at very late times, say, they must have
been very large at early times. This relation also indicates that
equally distributed fluctuations are typically of the size
$\sqrt{c_3}\sim \sqrt{\hbar H}$ and thus $(\Delta
p)/p=\sqrt{G^{pp}}/p\sim \sqrt{\hbar/H}$.

\subsubsection{Saturation}

Of particular interest is the case of coherent states which saturate
the uncertainty relations. For the harmonic oscillator, such states
are squeezed Gaussian states of the form
$\psi(q)=\exp(-z_1q^2+z_2q+z_3)$ with three complex numbers $z_i$ such
that ${\rm Re}z_1>0$. While ${\rm Re}z_3$ is fixed by normalization
and ${\rm Im}z_3$ is only a phase factor, $z_1=\alpha_1+i\beta_1$ and
$z_2=\alpha_2+i\beta_2$ determine the peak and fluctuations of the
state. One can easily see that
\begin{eqnarray}
 \langle\hat{q}\rangle&=& \frac{\alpha_2}{2\alpha_1}\,,\\
 \langle\hat{p}\rangle&=&
 \frac{\alpha_1\beta_2-\alpha_2\beta_1}{\alpha_1}\hbar\,,\\
 G^{qq} &=& \frac{1}{4\alpha_1}\,,\\
 G^{pp} &=& \alpha_1\hbar^2+\frac{\beta_1^2}{\alpha_1}\hbar^2\,,\\
 G^{qp} &=& -\frac{\beta_1}{2\alpha_1}\hbar
\end{eqnarray}
such that indeed
\begin{equation}
 G^{qq}G^{pp}- (G^{qp})^2= \frac{1}{4}\hbar^2\,.
\end{equation}
As is well known, these states even describe dynamical coherent states
of the harmonic oscillator, i.e.\ their form is preserved during
evolution. Saturation of the uncertainty relation leaves two free
second moments such as one spread parameter $G^{qq}$ and squeezing
$G^{qp}$. Unsqueezed states imply $\beta_1=0=G^{qp}$ and only
one free parameter specifies the width of the Gaussian.

Our system is different and no operator for $c$ (which would be an
analog of $\hat{q}$) exists. We have to work with exponentials
instead, and thus even kinematical coherent states change compared to
the harmonic oscillator. For dynamical coherent states the form must
anyway be different since we are not dealing with the harmonic
oscillator Hamiltonian.  Thanks to the solvability of our model we are
still able to determine properties of dynamical coherent states
explicitly. This provides an instructive example of how properties of
coherent states, and physical implications, can change when a system
is not closely related to a harmonic oscillator.

We thus look at saturation of our uncertainty relations
(\ref{uncertI}), (\ref{uncertII}) and (\ref{uncertIII}), first
removing the $\phi$-dependence. Subtracting (\ref{uncertIII}) divided by
four from (\ref{uncertII}) yields
\begin{equation} \label{satI}
 c_2^2-c_1^2-4c_5c_6=-\hbar^2AB\,.
\end{equation}
The coefficients of $e^{-2\phi}$ and $e^{2\phi}$ in (\ref{uncertII}) then
yield two more independent relations
\begin{eqnarray}
 (c_1-c_2)c_3-c_6^2 &=& \frac{1}{4}\hbar^2A^2\,,\label{satII}\\
 (c_1-c_2)c_4-c_5^2 &=& \frac{1}{4}\hbar^2B^2 \label{satIII}
\end{eqnarray}
in addition to (\ref{uncertI}) which becomes
\begin{equation}
 4c_3c_4-\frac{1}{4}(c_1+c_2)^2=\hbar^2H^2\,. \label{satIV}
\end{equation}
There are thus four equations for six variables, such that two remain
free as in the case of Gaussians. Note, however, that the reality
condition relates one of these, $c_1$, to the expectation values, or
the constants $A$, $B$ and $H$. Thus, only one combination of the
uncertainty parameters is free, which we can take as $c_1-c_2=2(\Delta
H)^2$, and saturated states are more restricted than for the harmonic
oscillator. (This is a consequence of the non-linear reality condition
which relates some fluctuations to expectation values.)

Without loss of generality we can assume $A=B$ for bounces or $A=-B$
for non-bouncing solutions since it simply amounts to choosing the
origin of time $\phi$ such that the bounce (or the transition through
$p=0$) occurs at $\phi=0$.  Subtracting (\ref{satII}) and
(\ref{satIII}) then gives
\begin{equation}\label{satdiff}
 (c_1-c_2)(c_3-c_4)=c_6^2-c_5^2
\end{equation}
which shows that $p$-fluctuations are the same before and after the
bounce, i.e.\ $c_3=c_4$, if and only if also $|c_5|=|c_6|$ (recall that
$c_1-c_2=2(\Delta H)^2$ cannot be zero). This case will be discussed
below.

Using $A=\pm B$, we can combine (\ref{satI}), (\ref{satII}) and
(\ref{satIII}) to obtain a further $A$-independent relation
\begin{equation}
 (c_1-c_2)(c_3+c_4)- (c_5\pm c_6)^2\pm\frac{1}{2}(c_2^2-c_1^2)=0\,.
\end{equation}
Solving for $c_3+c_4$ and combining it with (\ref{satdiff}) gives
\begin{eqnarray}
 c_3 &=& \frac{\pm c_5+c_6}{c_1-c_2}c_6\pm\frac{1}{4}(c_1+c_2)\,, \label{c3}\\
 c_4 &=& \frac{c_5\pm c_6}{c_1-c_2}c_5\pm\frac{1}{4}(c_1+c_2) \label{c4}
\end{eqnarray}
in terms of only $c_1$, $c_2$, $c_5$ and $c_6$. Using this in
(\ref{satIV}) and combining it with (\ref{satI}), we have
\begin{eqnarray}
 \hbar^2H^2+\frac{1}{4}(c_1+c_2)^2 &=& 4c_3c_4\nonumber\\
  &=& \pm 4\left(\frac{c_5\pm c_6}{c_1-c_2}\right)^2 c_5c_6\pm
 \frac{c_1+c_2}{c_1-c_2} (c_5\pm c_6)^2+
\frac{1}{4}(c_1+c_2)^2\nonumber\\
 &=& \pm \left(\frac{c_5\pm c_6}{c_1-c_2}\right)^2
(4c_5c_6+c_1^2-c_2^2)+ \frac{1}{4}(c_1+c_2)^2\nonumber\\
 &=& \hbar^2A^2\left(\frac{c_5\pm c_6}{c_1-c_2}\right)^2+
 \frac{1}{4}(c_1+c_2)^2
\end{eqnarray}
and thus (with $c_1-c_2=2(\Delta H)^2>0$)
\begin{equation} \label{c5c6}
 |c_5\pm c_6| = \frac{H}{A}(c_1-c_2)\,.
\end{equation}
This shows that it is impossible to have both $c_5$ and $c_6$ zero,
i.e.\ $G^{p,i(J-\bar{J})}\not=0$ and there are always correlations
between $p$ and $H$ which evolve in time. Together with
(\ref{satI}), (\ref{c3}) and (\ref{c4}), this last relation allows us
immediately to express all parameters in terms of only $c_1$ and
$c_2$, and all relations for saturation are solved.

We now ask whether it is possible to have a coherent state which
behaves semiclassically at one (late) time and has identical
fluctuations before and after the bounce. We thus focus on the case
$c_5=\pm c_6$.
This assumption allows us to solve (\ref{c5c6}) directly for $c_5$,
\begin{equation}
 |c_5|=|c_6|= \frac{H}{2A}(c_1-c_2)
\end{equation}
and to insert it in (\ref{c3}),
\begin{equation} \label{c3c4}
 c_3=c_4= \frac{H^2}{2A^2}(c_1-c_2)\pm\frac{1}{4}(c_1+c_2)\,.
\end{equation}
Consistency with (\ref{satIV}) implies
\begin{equation}
 4c_3^2=\hbar^2H^2+\frac{1}{4}(c_1+c_2)^2=
 \left(\frac{H^2}{A^2}(c_1-c_2)\pm \frac{1}{2}(c_1+c_2)\right)^2\,.
\end{equation}
Coherent states with identical fluctuations before and after the
bounce are thus possible if and only if
\begin{equation} \label{DeltaH4}
 \frac{H^4}{A^4}(c_1-c_2)^2\mp\frac{H^2}{A^2}(c_2^2-c_1^2)=
 \hbar^2H^2\,.
\end{equation}
Using $c_1-c_2=2(\Delta H)^2$ and $c_1+c_2=-2(\Delta H)^2+2c_1=
-2(\Delta H)^2-2H^2\pm 2A^2+\frac{1}{2}\hbar^2$ (imposing the reality
condition (\ref{realityAB})), we must then solve
\[
 \frac{H^2\mp A^2}{H^2}(\Delta H)^4\mp \frac{A^2(H^2\mp
A^2-\frac{1}{4}\hbar^2)}{H^2}(\Delta
H^2)-\frac{A^4}{H^2}\frac{\hbar^2}{4}=0
\]
giving
\begin{eqnarray}
 (\Delta H)^2 &=& \pm\frac{A^2(H^2\mp A^2-\frac{1}{4}\hbar^2)}{2(H^2\mp
A^2)} +\sigma \frac{\sqrt{A^4(H^2\mp A^2-\frac{1}{4}\hbar^2)^2+A^4(H^2\mp
A^2)\hbar^2}}{2(H^2\mp A^2)}\nonumber\\
 &=& \frac{\mp A^2((H^2\mp A^2-\frac{1}{4}\hbar^2)+\sigma |H^2\mp
A^2+\frac{1}{4}\hbar^2|)}{2(H^2\mp A^2)}\,.
\end{eqnarray}
%\frac{c_1(H^2+c_1-\hbar^2/4)}{2(-c_1+\hbar^2/4)}
% \pm\frac{H^2+c_1-\hbar^2/4}{-4c_1+\hbar^2}\sqrt{4c_1^2-4c_1\hbar^2+\hbar^4}
Here we distinguished the two roots of the quadratic equation
(\ref{DeltaH4}) by $\sigma=\pm 1$ since another $\pm$ has already been
used for the two cases $A=\pm B$.

Depending on the signs involved there are four possibilities to have
positive solutions for $(\Delta H)^2$:
\begin{enumerate}
 \item $A=B$, in which case there is a further distinction
\begin{enumerate} 
 \item $A^2<H^2+\frac{1}{4}\hbar^2$: Only $\sigma=1$
   is allowed, implying $(\Delta H)^2= A^2$.  
 \item $A^2>H^2+\frac{1}{4}\hbar^2$: Both signs for $\sigma$ are allowed, 
\begin{enumerate} 
 \item $\sigma=-1$ implies $(\Delta H)^2= A^2$ as above;
 \item $\sigma=1$ implies
\begin{equation} \label{1bii}
 (\Delta H)^2= \frac{A^2}{A^2-H^2}\frac{\hbar^2}{4}
\end{equation}
\end{enumerate}
\end{enumerate}
 \item $A=-B$, which allows only one choice of signs for a positive
\begin{equation}
 (\Delta H)^2= \frac{A^2}{H^2+A^2}\frac{\hbar^2}{4}\,.
\end{equation}
\end{enumerate}
There are two cases where $(\Delta H)^2 = A^2$ which can satisfy the
basic condition $(\Delta H)^2\ll H^2$ only if $A^2\ll H^2$. Thus $c_1$
must be of the order $H^2$. This can only happen if the bounce scale
$p(0)=A-\frac{1}{2}\hbar$ is small compared to the total energy, i.e.\
the universe enters the deep Planck regime during the bounce.
However, large $c_1$ (and thus large $c_2$ since $c_1-c_2$ must remain
small) of the order $H$ implies, using (\ref{c3c4}), that $c_3$ is of
the order $H$, too, and $\Delta p$ is not small compared to $p$ (in
fact, not even smaller). This case does not give rise to semiclassical
states at any time.

The case $A=-B$ allows small $\Delta H$. However, as we already saw,
the reality condition allows such solutions only if
$c_1=-H^2-A^2+\frac{1}{4}\hbar^2$ is large and negative. For states
saturating the uncertainty relations, this implies that $c_3$ is of
the same size as $-c_1$ and thus too large for the state to be
semiclassical at any time.

For the last possibility (\ref{1bii}), $(\Delta H)^2$ can be small
compared to $H^2$. For instance, if $c_1$ is of the order $A\hbar$,
which is allowed for semiclassical states, we have $(\Delta H)/H\sim
\sqrt{A\hbar}/H\sim \sqrt{\hbar/H}$. Moreover, $(\Delta p)/p\sim
\sqrt{\hbar/A}$ such that this final possibility does allow
semiclassical states with equal spread before and after the
bounce. This confirms the earlier indication that equally distributed
fluctuations typically satisfy $c_3\sim\hbar H$. Although there is a
factor of $\hbar$, fluctuations are rather large due to the factor of
$H$ which, for a universe with large matter energy, is a large
number. It is possible to have smaller fluctuations which are not
magnified by the matter energy, but only at one side of the bounce and
at the expense of having much larger fluctuations at the other side of
the bounce.

\begin{figure}
\begin{center}
  \includegraphics[height=.25\textheight]{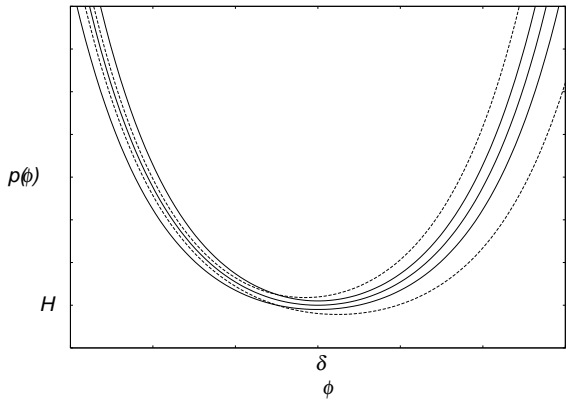} \caption{Two
  bouncing solutions for the expectation value of $\hat{p}$ and the 
  spread around
  it. Generic states have different spread before and after the bounce
  (dashed lines), while unsqueezed Gaussian initial states lead to
  solutions which are symmetric around the bounce not only in their
  expectation values but also in spreads (solid lines).
  \label{EffBounce}}
\end{center}
\end{figure}

There is an easier way to have symmetric fluctuations if one does not
require that all uncertainty relations be saturated. One can argue
that (\ref{uncertII}) is of primary interest since it determines the
fluctuations of $p$ and $H$, and that only this relation should be
saturated. If this is done, symmetric fluctuations before and after
the bounce are easily allowed. However, this can only be put in by
assumption and not be inferred from conditions at one time after the
bounce: only one of the relevant parameters $c_3$ or $c_4$ is
controlled by the uncertainty relation at one late or early time
$\phi$ while the other one would be suppressed exponentially by
$e^{-2|\phi|}$. Symmetric fluctuations before and after the bounce can
thus not be proven but only be assumed for coherent states of this
system. Generically, even a universe which is semiclassical at late
times can have been highly quantum before the big bang. Examples for
one symmetric and the generic non-symmetric bounces are shown in
Fig.~\ref{EffBounce}.

\subsubsection{Gaussian states}

This seems to be in conflict with recent numerical results
\cite{QuantumBigBang,APS,APSII} where the $p$-fluctuations turned out
to be very close to each other before and after the bounce. Indeed,
this gave rise to statements to the extent that the universe was as
semiclassical before the big bang as it is now. To resolve the
apparent contradiction we have to look at {\em unsqueezed Gaussian
states} as they were used by construction in those numerical
simulations (explicitly removing phase factors so as to de-squeeze
states), but are not covered by our preceding coherent state analysis.
As emphasized before Gaussian states do not saturate the uncertainty
relations in this system. They are thus not coherent, and not
distinguished as they would be for usual quantum mechanical systems.
Nevertheless, their properties are quite interesting.

Let us thus assume that we have a state of the form
$\psi(p)=N\exp(-z_1p^2+z_2p)$, supported on integer $p$, with ${\rm
Re}z_1>0$. Although these states have the same form as in standard
quantum mechanics, the representation of basic operators is
different. For instance, we have
\begin{eqnarray}
 \langle\hat{p}\rangle &=& \sum_p p|\psi(p)|^2\,,\\
 \langle\hat{J}\rangle &=& \sum_p p\bar{\psi}(p)\psi(p-\hbar)
\end{eqnarray}
where we sum over integers and the shift by $\hbar$ arises from the
action of $\widehat{e^{ic}}$ which is a shift operator in $p$. For
states which are nearly constant on the discrete scale of $p$, the
expressions can be seen as Riemann sums and approximated by Gaussian
integrals which one can compute explicitly.

We thus obtain
\begin{eqnarray}
 p &\approx& \frac{\alpha_2}{2\alpha_1}\,,\\
 J &\approx& \frac{\alpha_2+\alpha_1\hbar+i\beta_1\hbar}{2\alpha_1}
 \exp(-(\alpha_1^2+ \beta_1^2)\hbar^2/2\alpha_1
 -i(\beta_2-\alpha_2\beta_1/\alpha_1)\hbar)\,,\\
 G^{pp} &\approx& \frac{1}{4\alpha_1}\,,\\
 G^{pJ} &\approx& \!\!\!\!\left(\frac{1}{4\alpha_1}-
   \frac{\beta_1^2}{4\alpha_1^2}\hbar^2+
   \frac{i\beta_1\hbar}{4\alpha_1}\left(\hbar+\frac{\alpha_2}{\alpha_1}\right)\right)
 \exp(-(\alpha_1^2+\beta_1^2)\hbar^2/2\alpha_1-i\hbar(\beta_2-\alpha_2\beta_1/\alpha_1))\,,\\
 G^{JJ} &\approx&\!\!\!\! \left(\exp(-(\alpha_1^2+\beta_1^2)\hbar^2/\alpha_1)
   \left(\frac{1}{4\alpha_1}+\frac{\alpha_2^2}{4\alpha_1^2}-
     \frac{\beta_1^2}{\alpha_1^2}\hbar^2+
     \frac{\alpha_2\hbar}{2\alpha_1}+
     i\frac{\alpha_2\beta_1\hbar}{\alpha_1^2}\right)\right.
   -\frac{\alpha_2^2}{4\alpha_1^2}- \frac{\hbar^2}{4}\\
&& \left.
     +\frac{\beta_1^2}{4\alpha_1^2}\hbar^2-
     \frac{\alpha_2\hbar}{2\alpha_1}-
     i\frac{\beta_1\hbar}{2\alpha_1}(\hbar+\alpha_2/\alpha_1)\right)
 \exp(-(\alpha_1^2+\beta_1^2)\hbar^2/\alpha_1-
 2i(\beta_2-\alpha_2\beta_1/\alpha_1)\hbar)\,,\nonumber\\
 &\approx& \overline{G^{J\bar{J}}}\nonumber\\
 G^{J\bar{J}} &\approx& \frac{1}{4\alpha_1}+ \frac{\hbar^2}{2}+
 \frac{\alpha_2\hbar}{2\alpha_1}+ \frac{\alpha_2^2}{4\alpha_1^2}-
 \left(\frac{\alpha_2^2}{4\alpha_1^2}+
   \frac{\alpha_2\hbar}{2\alpha_1}+ \frac{\hbar^2}{4}+
   \frac{\beta_1^2\hbar^2}{4\alpha_1^2}\right)
 \exp(-(\alpha_1^2+\beta_1^2)\hbar^2/\alpha_1)\,.
\end{eqnarray}
These expressions involving $J$ are much more messy than the
corresponding ones for Gaussian states in standard quantum
mechanics. This is a consequence of the fact that Gaussian states for
the system considered here are not natural at all.

Nevertheless, the expressions simplify somewhat if one assumes that
the state is unsqueezed, $\beta_1=0$. Moreover, for semiclassical
states we can use $p\gg\Delta p\gg \hbar$ which implies
$\alpha_2/\alpha_1\gg \alpha_1^{-1/2}\gg\hbar$. Then, the leading
order contribution to the energy fluctuations, derived using
\begin{eqnarray*}
 {\rm Re}\,G^{JJ} &=&
\left(\left(\frac{1}{4\alpha_1}+\frac{\alpha_2^2}{4\alpha_1^2}+
\frac{\alpha_2}{2\alpha_1}\hbar\right) e^{-\alpha_1\hbar^2}-
\frac{\alpha_2^2}{4\alpha_1^2}-\frac{\hbar^2}{4}-
\frac{\alpha_2}{2\alpha_1}\hbar\right)
e^{-\alpha_1\hbar^2}\cos(2\beta_2\hbar)\\
 &\approx& -\frac{\alpha_2^2}{4\alpha_1}\hbar^2\cos(2\beta_2\hbar)
\end{eqnarray*}
and
\begin{eqnarray*}
 G^{J\bar{J}} &=&
\frac{1}{4\alpha_1}+\frac{\hbar^2}{2}+\frac{\alpha_2}{2\alpha_1}\hbar+
\frac{\alpha_2^2}{4\alpha_1^2}- \left(\frac{\alpha_2^2}{4\alpha_1^2}+
\frac{\alpha_2}{2\alpha_1}\hbar+ \frac{\hbar^2}{4}\right)
e^{-\alpha_1\hbar^2}\\
 &\approx& \frac{\alpha_2^2}{4\alpha_1}\hbar^2\,,
\end{eqnarray*}
is
\begin{equation}
 G^{HH}=-\frac{1}{4}(G^{JJ}+G^{\bar{J}\bar{J}})+\frac{1}{2}G^{J\bar{J}}
\approx \frac{\alpha_2^2}{4\alpha_1}\hbar^2
 \cos^2(\beta_2\hbar)
\end{equation}
and with $H={\rm Im}J\approx -\alpha_2\sin(\beta_2\hbar)/2\alpha_1$ we
have
\begin{equation}
 \frac{G^{HH}}{H^2} \approx \alpha_1\hbar^2\cot^2(\beta_2\hbar)
\end{equation}
whose left hand side is constant throughout the whole evolution. At
late and early times $\beta_2$ becomes small such that
$\alpha_1\propto\cot^{-2}(\beta_2\hbar)\approx \sin^2(\beta_2\hbar)$.
This relates $\alpha_1$ which determines the $p$-fluctuation to
$\beta_2$ which determines the peak position of the wave packet. Thus,
for unsqueezed Gaussians we prove that the fluctuations before and
after the bounce are the same at times where $\sin^2(\beta_2\hbar)$
takes the same value. Since the expectation value solutions
(\ref{psol}) are symmetric around the bounce point for any state,
{\em fluctuations of unsqueezed Gaussian states are shown to be symmetric
around the bounce.}

This reconciles our calculations with the numerical calculations of
\cite{APS,APSII} and reinforces their validity. However, it also
demonstrates that the result of identical fluctuations before and
after the bounce is a consequence not of the generic dynamics of
semiclassical states, but relies on the assumption that states are
unsqueezed Gaussians. It is then not very surprising to find symmetric
spreads since there is a single parameter determining the state, other
than its expectation values. The fact that $G^{HH}$ is constant then
implies directly that there is a fixed relation between this
parameter, $\alpha_1$, and the peak position. Since expectation values
are symmetric around the bounce for any solution, the spread must
satisfy the same symmetry in this restricted case.  As discussed
before, there is no intrinsic basis in this model to restrict states
to such a form. They do not saturate the uncertainty relations, and
even if the Gaussian form is assumed but general squeezing is allowed
fluctuations before and after the bounce become independent of each
other. There is then an additional parameter $\beta_1$ and only a
certain function of spread $\alpha_1$ and squeezing $\beta_1$ is fixed
at the bounce-reflected point. This does not suffice to fix the spread
to be symmetric. The precise relation follows by estimating $G^{HH}$
as above, now keeping $\beta_1\not=0$. Then, $G^{HH}/H^2={\rm const}$
with
\[
 G^{HH}\approx
\frac{\alpha_2^2}{4\alpha_1^3}(\alpha_1^2+\beta_1^2)\hbar^2
\cos^2((\beta_2-\alpha_2\beta_1/\alpha_1)\hbar)
\]
and
\[
 H\approx
-\frac{\alpha_2}{2\alpha_1}\sin((\beta_2-\alpha_2\beta_1/\alpha_1)\hbar)+
\frac{\beta_1}{2\alpha_1}\hbar
\cos((\beta_2-\alpha_2\beta_1/\alpha_1)\hbar)
\]
provides the relation between $\alpha_1$ and $\beta_1$ in terms of the
expectation values.

\subsubsection{The role of superposed branches}
\label{role}

\begin{figure}
\begin{center}
  \includegraphics[height=.35\textheight]{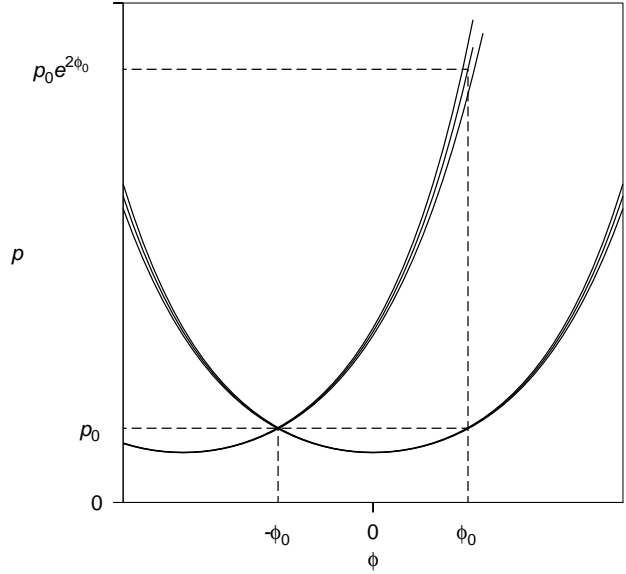} \caption{Sketch
  of two wave functions with the same peak position $p_0$ at
  $-\phi_0$, but one being collapsing the other expanding. After the
  bounce of the first solution at $\phi=0$, the wave packets deviate
  strongly at $\phi_0$. Both wave packets are illustrated by their
  expectation values and spreads, assumed to be symmetric around the
  bounce for the sake of the argument.
\label{Superposed}}
\end{center}
\end{figure}

One could suspect that non-symmetric fluctuations are a consequence of
small admixtures of superposed expanding branches on an initial
contracting one, while superpositions of energy eigenstates of
definite sign might always have symmetric fluctuations.  As
illustrated in Fig.~\ref{Superposed}, we choose two states
$|\psi_1\rangle$ and $|\psi_2\rangle$ at $\phi=-\phi_0$ which are
peaked at the same value $p_0$ of $p$ with the same $p$-fluctuations.
If the first state is expanding,
$\langle\psi_1|\hat{H}|\psi_1\rangle=:H>0$, while the second one is
contracting, $\langle\psi_2|\hat{H}|\psi_2\rangle=
-\langle\psi_1|\hat{H}|\psi_1\rangle<0$, they are sharply peaked at
different $H$, and thus $\langle\psi_1|\psi_2\rangle\ll1$. Any
superposition
$|\psi\rangle=(|\psi_1\rangle+\alpha|\psi_2\rangle)/\sqrt{1+|\alpha|^2}$,
where $|\psi_2\rangle$ presents an admixture of a negative-$H$ state,
is then sharply peaked at $p_0$.  At $\phi=\phi_0$, expectation values
of $\hat{p}$ in the two states of the superpositon have evolved away
from each other and the combined spread $\Delta p$ can be much larger
than individual spreads $\Delta_1p$ and $\Delta_2p$ measured in
$|\psi_1\rangle$ and $|\psi_2\rangle$, respectively.

This can be analyzed more quantitatively: For simplicity, we assume
that the $H$-fluctuations of the two states are nearly equal,
$\Delta_1H\approx\Delta_2H$.  In the state $|\psi\rangle$, we have
\begin{eqnarray*}
 (\Delta H)^2&=& \langle\psi|\hat{H}^2|\psi\rangle-
 \langle\psi|\hat{H}|\psi\rangle^2 \approx \frac{1}{1+|\alpha|^2}
 (\langle\hat{H}^2\rangle_1+ |\alpha|^2\langle\hat{H}^2\rangle_2)-
 H^2\left(\frac{1-|\alpha|^2}{1+|\alpha|^2}\right)^2\\
&=& \frac{1}{(1+|\alpha|^2)^2}((\Delta_1H)^2(1+|\alpha|^2)+
(\Delta_2 H)^2|\alpha|^2 (1+|\alpha|^2)+ 4|\alpha|^2H^2)
\end{eqnarray*}
where subscripts of 1 and 2 at right brackets and $\Delta$ indicate
which state is used for the expectation values. Now assuming
$(\Delta_1 H)^2\approx (\Delta_2H)^2$, we obtain
\begin{equation}
\frac{(\Delta H)^2}{H^2} \approx 
\frac{(\Delta_1H)^2}{H^2}+
 4\frac{|\alpha^2|}{(1+|\alpha|^2)^2}\,.
\end{equation}
Thus, the admixture changes the $H$-fluctuations only slightly and
preserves $\Delta H\ll H$ if $|\alpha|$ is sufficiently small. It
would thus be allowed in our approximation, although not in a
superposition of only positive energy eigenstates. We will now see how
the admixture can influence spreads of $\hat{p}$ before and after the
bounce.

While $(\Delta H)/H$ is constant in time, $(\Delta p)/p$ changes and the
initial value which by construction is close to $(\Delta_1p)/p$ is not
conserved. Using the behavior of exact solutions of expectation
values, one can compute $p$-fluctuations at the bounce-reflected point
of the initial state. At this time, the two states will, in the
$p$-representation, have evolved away from each other since one state
corresponds to an expanding branch and the other to a collapsing
one. We have
\begin{eqnarray*}
 (\Delta p)^2 &=& \frac{1}{1+|\alpha|^2}(\langle\hat{p}^2\rangle_1+
|\alpha|^2\langle\hat{p}^2\rangle_2)-
\left(\frac{\langle\hat{p}\rangle_1+
|\alpha|^2\langle\hat{p}\rangle_2}{1+|\alpha|^2}\right)^2\\ &=&
\frac{1}{(1+|\alpha|^2)^2}
(((\Delta_1p)^2+|\alpha|^2(\Delta_2p)^2)(1+|\alpha|^2)+
|\alpha|^2\langle\hat{p}\rangle_1^2+
|\alpha|^2\langle\hat{p}\rangle_2^2-
2|\alpha|^2\langle\hat{p}\rangle_1 \langle\hat{p}\rangle_2)\\ &=&
(\Delta_1p)^2 \frac{1+|\alpha|^2
\langle\hat{p}\rangle_2^2/\langle\hat{p}\rangle_1^2}{1+|\alpha|^2}+
\frac{|\alpha|^2}{(1+|\alpha|^2)^2}
(\langle\hat{p}\rangle_1-\langle\hat{p}\rangle_2)^2\,. \label{pspread}
\end{eqnarray*}
In the last step we used the fact that by construction
$|\psi_2\rangle$ is in the expanding branch at all times considered,
thus $(\Delta p)/p\approx {\rm const}$ away from the bounce and
$(\Delta_2p)^2\approx
\langle\hat{p}\rangle_2^2(\Delta_1p)^2/\langle\hat{p}\rangle_1^2>
(\Delta_1p)^2$. For instance, at the bounce-reflected $\phi_0$ of
$|\psi_1\rangle$ with large $\phi_0$, we have
\[
 \langle\hat{p}\rangle_2(\phi_0)\approx p_0e^{2\phi_0}\approx
 \langle\hat{p}\rangle_1(\phi_0)
(2p_0/H)^2
\]
and thus
\begin{equation}
 \frac{(\Delta p)(\phi_0)^2}{p(\phi_0)^2} = 
 \frac{(\Delta_1p)(\phi_0)^2}{p(\phi_0)^2}\frac{1+4|\alpha|^2
p_0^2/H^2}{1+|\alpha|^2}+ \frac{16|\alpha|^2}{(1+|\alpha|^2)^2}
\left(\frac{p_0}{H}\right)^4 
\frac{\langle\hat{p}\rangle_1(\phi_0)^2}{p(\phi_0)^2}\,.
\end{equation}
Due to the large factor $p_0^2/H^2\gg 1$ for an initial state peaked
at large volume, the $p$-fluctuation have grown much more than the
$H$-fluctuation if we pass through the bounce. 

To verify that this implies non-symmetric fluctuations we have to
compute $\Delta p$ at the bounce-reflected point of
$\langle\hat{p}\rangle$, which is not $\phi_0$ due to the contribution
from $|\psi_2\rangle$. From 
\[
 \langle\hat{p}\rangle=
 \frac{\langle\hat{p}\rangle_1+ 
|\alpha|^2\langle\hat{p}\rangle_2}{1+|\alpha|^2} \approx
 \langle\hat{p}\rangle_1+ |\alpha|^2\langle\hat{p}\rangle_2
\]
for small $|\alpha|$, we have
\[
 \langle\hat{p}\rangle(\phi)\approx H\cosh\phi+
 \frac{1}{2}|\alpha|^2He^{\phi+2\phi_0}
\]
which has its minimum at $\phi_{\rm bounce}=
-\log\sqrt{1+|\alpha|^2e^{2\phi_0}}$. The bounce-reflected point for
$\langle\hat{p}\rangle$ is thus at $\phi_{\rm reflected}=\phi_{\rm
  bounce}+(\phi_{\rm bounce}-(-\phi_0))=
\phi_0-\log(1+|\alpha|^2e^{2\phi_0})$. Evaluating (\ref{pspread}) at
this point, for simplicity assuming $|\alpha|^2e^{2\phi_0}\gg1$
which presents a characteristic example for an admixture at large initial
volume, gives
\[
 \frac{(\Delta p)(\phi_{\rm reflected})^2}{p(\phi_{\rm reflected})^2}
 \approx \frac{(\Delta
   p)(-\phi_0)^2}{p(-\phi_0)^2}
 \frac{1}{|\alpha|^2} \frac{|\alpha|^4+|\alpha|^{-4}
   e^{-2\phi_0}}{|\alpha|^4+2e^{-2\phi_0}+|\alpha|^{-4} e^{-2\phi_0}}+
 \frac{1}{|\alpha|^2(1+|\alpha|^2)^2} \,.
\]
in terms of the initial values at $-\phi_0$.  The additional inverse
powers of the small $|\alpha|$ compared to the spread of $H$ show the
growth of $p$-fluctuations after the bounce, demonstrating that the
superposition will not have symmetric spread.

One could thus suspect that such an admixture, which would not violate
our condition $\Delta H\ll H$ if $\alpha$ is sufficiently small, could
be the reason for unequal dispersions before and after the bounce,
while solutions of exactly positive $H$ would have symmetric
dispersions as explicitly shown for unsqueezed Gaussians. This
conjecture cannot be true, however, because the solutions we studied
earlier only refer to expectation values and dispersions and, due to
the decoupling in our solvable model, are completely independent of
higher moments. The preceding construction of the admixture does
provide states with suitable initial dispersions and expectation
values, but the specific states
$|\psi\rangle=(|\psi_1\rangle+\alpha|\psi_2\rangle)/\sqrt{1+|\alpha|^2}$
also have fixed higher moments. There are many other states having the
same expectation values and dispersions but different higher moments,
not corresponding to what one obtains from an admixture of a negative
energy state. Such states allow for positivity as well as
non-symmetric dispersions, which thus cannot be an artifact of a
negative energy admixture. In short, the calculation confirms the
intuitive expectation that an admixture does give non-symmetric
spreads even if each state in the superposition has symmetric
spreads. But it does not show the converse, namely that non-symmetric
fluctuations could only be caused by an admixture of a negative energy
state.

\section{Conclusions}
\label{Concl}

A solvable model such as the one discussed here allows a detailed
analysis of dynamical coherent states which would otherwise be
difficult to handle. This provides valuable information for quantum
cosmology, just as the harmonic oscillator does traditionally for
quantum optics (see, e.g., \cite{SqueezedStates}). Compared to the
harmonic oscillator, our system shows several new properties with
implications for cosmology.

\subsection{Spreading of states}

Although the spreads are not constant for solutions to our system,
ratios such as $(\Delta p)/p$ are nearly constant in each pre- and
post-bounce branch. Nevertheless, this ratio can, and in general does,
take different values in both branches.  Dynamical coherent states
which exactly saturate the uncertainty relations and have spreads
symmetric around the bounce point do exist but are not
generic. Nevertheless, they are distinguished in a certain sense and
can thus be seen as analogs of the harmonic oscillator ground state
(although no unique symmetric state exists). Indeed, for such states
the scale of fluctuations, $(\Delta p)/p\approx \sqrt{\hbar/H}$, is
determined more sharply than without the symmetry assumption.

Dynamical coherent states for the loop quantization are not Gaussians
which turn out to have different and rather special
properties. Nevertheless, Gaussian states can also be analyzed
straightforwardly in this setting, with results being ultimately in
agreement with recent numerical investigations. Interestingly, {\em
unsqueezed} Gaussian states do have identical spread before and after
the bounce. They cannot be coherent but may well serve as a
special version of semiclassical states. This illustrates how
differently coherent states in a new system can behave from those
well-known for the harmonic oscillator. It also shows that a dynamical
analysis of coherent and semiclassical states is always necessary even
to select suitable initial states to be evolved. While dynamical
coherent states are difficult to describe in most systems, solvable
models make this possible which is now also available for cosmological
systems.

As demonstrated, the techniques of \cite{EffAc} reviewed in
Sec.~\ref{Method} provide an efficient way to derive coherent state
properties. Moreover, they allow one to see how properties can change
under perturbations away from the solvable model, such as by including
a scalar potential. Although this has not been dealt with in the
present paper, our detailed analysis of dynamical coherent states of
the solvable system provides crucial information for the zeroth order
of such a perturbation theory.

\subsection{Applications to cosmology}

There is only scant information on properties of the present state of
our universe other than that it is, to a high degree,
semiclassical. But this does not tell us which semiclassical or even
coherent state describes it best. From the harmonic oscillator or free
quantum field theories one is used to unsqueezed Gaussians to
represent the vacuum state. But the form of such states depends on the
system being looked at, and quantum cosmology is not close to either
the harmonic oscillator or free quantum field theory. Moreover, there
is no obvious vacuum state for quantum cosmology, and even if a
Hamiltonian should allow a ground state, it is unlikely to describe a
universe able to expand to large volume. A dedicated analysis as done
here shows which coherent states are available and what their generic
properties are. These properties must be taken into account for robust
statements about quantum cosmological systems.

For instance, we have seen that squeezing must be allowed for generic
states, which crucially changes properties such as the symmetry of
spreads before and after a bounce. In fact, squeezing of semiclassical
states often plays a large role for decoherence or the transition to
classical behavior (e.g.\ in the context of inflation
\cite{AlbSqueeze,CosmoDecoh,QuantClassCosmo}). Thus, a semiclassical
state at large volume of a cosmological model should indeed be assumed
to be highly squeezed. Lacking additional input, robust cosmological
conclusions can be drawn only with reference to {\em generic} coherent
states. Then, no strong restrictions on fluctuations of a state before
the bounce are justified. For all we know, it could have been coherent
but with large quantum fluctuations.

\subsection{Effective equations and the possibility not to bounce}

Although we have proven that the solvable quantum system is {\em
exactly} described by the effective Hamiltonian
\[
 H=\langle\hat{H}\rangle=\frac{1}{2i}(J-\bar{J})=p\sin c
\]
(determining the same equations of motion for $p$ and $J$ as $\hat{H}$
does for $\langle\hat{p}\rangle$ and $\langle\hat{J}\rangle$)
it is possible, depending on the initial state, that the system does
not bounce but reaches $p=0$ in finite proper time. This looks
contradictory at first sight since such an effective Hamiltonian
implies the effective constraint equation
\[
 H^2=p^2\sin^2c \propto p_{\phi}^2\,.
\]
With $p_{\phi}$ constant and $|\sin c|\leq 1$ there must be a non-zero
lower bound for $p$, the bounce scale.

However, while the effective Hamiltonian of our linear system does not
receive corrections from quantum back-reaction of fluctuations and
other quantum variables, the reality condition
$\hat{J}\hat{J}^{\dagger}=\hat{p}^2$ is non-linear. Classically, this
condition implies that $c$ is real and thus $|\sin c|\leq 1$. But the
reality condition (\ref{reality}) now receives quantum corrections of
second order in quantum variables which, for suitable states, can
remove the bounce. A state can then enter the ``classically forbidden
region'' where $|\sin c|>1$ while still respecting the quantum reality
condition.

As shown explicitly, zero volume is reached when the parameter $c_1$
is negative and large. While this can be achieved respecting the
spositivity condition $\Delta H\ll H$, such a state would never be
semiclassical. Thus, {\em any state which is semiclassical at one time
will give rise to a bounce.} These are all relevant states since a
boundary condition for modelling our universe, however distantly, is
always that there is at least one large volume regime in which the
state is semiclassical. But the possibility of states which do not
bounce demonstrates the non-triviality of the result. Simply replacing
$\dot{a}^2$ in the Friedmann equation by a bounded function is not
enough; any such replacement would have to be followed up by a
coherent state analysis which is much more non-trivial than an
analysis of the resulting ``effective'' Friedmann equation obtained by
the naive replacement. Loop quantum cosmology with a free scalar
passes this more stringent test and thus provides the first example in
loop quantum gravity where complete effective equations have been
computed.

We end by repeating that any physical statements derived from a single
model have to be confirmed by a perturbation analysis around the
model. This is feasible in our case, as it is for perturbations around
any solvable model, but still requires detailed work which is now in
progress. Only such an analysis could justify the transfer of results
from single models to our own universe. It may well be that this
removes the bounce through back-reaction of quantum variables
$G^{a,n}$ on the expectation values. In particular, it is then
conceivable that a state starts out perfectly semiclassically at large
volume where its expected volume collapses, evolves for a long time to
small volume and all along picks up corrections from quantum
back-reaction. Since also quantum variables evolve, it cannot be ruled
out without further analysis that the analog of $c_1$ does become
negative and large close to the would-be bounce. If this happens, the
bounce is avoided for the self-interacting state even if it starts out
semiclassically by all possible conditions one could pose. This is
only one possibility out of many which can only be ruled out by
performing a comprehensive perturbation analysis for which the results
of this paper present the zeroth order basis.

\section*{Acknowledgements}

The author thanks Hans Kastrup and Jurek Lewandowski for discussions.
This work was supported in part by NSF grants PHY99-07949 and
PHY05-54771.

%\bibliographystyle{../preprint}
%\bibliography{../Bib/QuantGra}

\end{document}